\documentclass[useAMS,usenatbib]{mn2e}

\usepackage{graphicx}

\newcommand{\sciHI}{H\,\small{I}}
\newcommand{\Tor}{{\sc torus}}

\title[Simulated Galactic \sciHI]{A Synthetic 21-cm Galactic Plane Survey of an SPH 
Galaxy Simulation}
\author[K.~A.~Douglas et al.]
  {Kevin A.~Douglas$^1$,\thanks{email: douglas@astro.ex.ac.uk} David M.~Acreman$^1$, 
  Clare L.~Dobbs$^{1,2,3}$, and 
\newauthor  Christopher M.~Brunt$^1$ \\
$^1$ School of Physics, University of Exeter, Stocker Road, Exeter, UK  EX4 4QL \\
$^2$ Max-Planck-Institut f\"ur extraterrestrische Physik, Giessenbachstra\ss{}e, D-85748 Garching, Germany \\
$^3$ Universitats-Sternwarte M\"unchen, Scheinerstra\ss{}e 1, D-81679 M\"unchen, Germany } 

\begin{document}


\pagerange{\pageref{firstpage}--\pageref{lastpage}} \pubyear{2010}

\maketitle

\label{firstpage}

\begin{abstract}
We have created synthetic neutral hydrogen ({\sciHI}) Galactic Plane Survey data cubes covering
$90^\circ \le \ell \le 180^\circ$,
using a model spiral galaxy from SPH simulations and the radiative transfer code {\Tor}.  The density,
temperature and other physical parameters are fed from the SPH simulation into {\Tor}, where the
{\sciHI} emissivity and opacity are calculated before the 21-cm line emission profile is determined.
Our main focus is the observation of Outer Galaxy `Perseus Arm' {\sciHI}, with a view to tracing atomic
gas as it encounters shock motions as it enters a spiral arm interface, an early step in the formation
of molecular clouds.  The observation of {\sciHI} self-absorption features at these shock sites (in both 
real observations and our synthetic data) allows us to investigate further the connection between cold atomic
gas and the onset of molecular cloud formation.
\end{abstract}


\section{Introduction}

The formation of molecular clouds out of cold, dense atomic gas is a key process in
the cyclic evolution of the interstellar medium.  
Above column densities of $N_H \approx 10^{20}$ cm$^{-2}$, atomic hydrogen gas transitions 
to molecular H$_2$, a critical step towards the formation of stellar systems \citep{dix98}.  For the molecular
fraction of the gas to remain stable depends largely on the ability of the molecular gas to
self-shield against dissociating photons from the local radiation field.  Moreover, the
importance of dust grains in H$_2$ formation requires their presence in these transition interfaces.

The 21-cm neutral hydrogen line is the main tracer of interstellar structure.
The ubiquitous nature of atomic hydrogen in the Galaxy was demonstrated in early surveys \citep{Oort58}, and
more recent surveys imaging the {\sciHI} distribution at moderately high resolution (parsec scales at distances of
kpc) have steadily improved upon that view.  Almost all of the {\sciHI} in the Galaxy's disc has
been sampled by the International Galactic Plane Survey (IGPS; \citet{vgps06,sgps05,tayetal03}).  Beyond 
the Galactic Plane, all-sky mapping efforts of the Southern (GASS; \citet{gass09}) and Northern skies (EBHIS;
\citet{ebhis09}) have
improved vastly on the previous generation's all-sky {\sciHI} maps.  Galactic 21 cm surveys at Arecibo 
with its ALFA instrument \citep{galfapaper,peek09,vanL09,putmanetal09} are revealing {\sciHI} structures in our
Galactic neighbourhood with unprecedented sensitivity.

Observational evidence of the interface between atomic and molecular gas relies on a good H$_2$ tracer.
The $J=1-0$ rotational transition of CO is the most common surrogate of molecular hydrogen, yet there are
conditions under which its efficacy as a reliable H$_2$ tracer is undermined \citep{dou07,kla05,dob07,pel09}.  
For example, in photon-dominated
regions, the CO molecule is largely dissociated, while H$_2$ self-shielding allows for a significant amount
of molecular gas to remain.  In very cold, dense molecular cores, CO will freeze onto dust grains along
with other larger molecules, and so the H$_2$ content in such cores would be underestimated by traditional
CO observations.  In addition to the above scenarios, which assume a molecular cloud has already formed, a
newly forming molecular cloud must have an epoch wherein a significant quantity of H$_2$ has formed before
an appreciable (detectable) amount of CO has been reached \citep{Ber04}, and so the evolution of protostellar
objects may be already underway in molecular clouds before the onset of critical CO densities.  
OH is another molecule which has been used to probe the H-H$_2$ transition region, with some success \citep{Mag90,Wan93}.
Neutral and ionised carbon atoms also have potential to be good tracers of diffuse, pre-CO molecular gas
\citep{oka05}.

In order for appreciable H$_2$ to form from a largely atomic hydrogen reservoir, the gas 
must be cooled to temperatures on the
order of 20 K, among other favourable conditions (sufficient number densities, H$_2$ formation sites).
Models of gas flowing into the Perseus spiral arm \citep{Roberts1972} predict compression of the {\sciHI}
gas at the interface of a ``shock ridge."  Though his model was isothermal, such conditions are favourable for 
strong cooling of {\sciHI}, leading to self-absorption of 21-cm
emission, and possibly enhanced H$_2$ production.  Hence {\sciHI} self-absorption (HISA) observations
may help to trace the earliest stages of molecular cloud formation, in regions where CO is not a workable tracer
for molecular hydrogen.  Widespread HISA has been observed in the Outer Galaxy \citep{Gibsonetal2005}, and its
association with CO is anything but straightforward.  HISA and CO emission are often found to be coincident
spatially and at the same radial velocities, but the correspondence is complicated.  There are regions
where only one is present, which may indicate differences in the evolutionary state of these clouds.   Studies of
HISA and OH emission have found the OH emission to be more localised to the HISA \citep{Kav06,Li03}, compared to 
CO.  As an extra complication, the
detection of HISA requires background {\sciHI} emission at the same velocity as the absorbing gas, which may not be 
available along many lines of sight.

In addition to analytical models, computational efforts have investigated this scenario in detail.  
\citet{DobbsGloverClarkKlessen2008} performed galaxy-scale SPH simulations that trace the dynamics of interstellar
clouds as they encounter shocks in spiral arms.  To enhance resolution in the spiral arms, their model traces only
those SPH particles in a galactocentric annulus between $R_g = 5$ kpc out to $R_g = 10$ kpc.
In addition to strong cooling of the atomic gas, molecular hydrogen
forms in the vicinity of the shocks.  Smaller clouds aggregrate into larger ones within the spiral shock, where
the molecular fraction is $\ge 0.5$.  As the clumps leave the spiral arm they are sheared into smaller spurs
and lose most of their molecular gas.

The link between these models and observations can be strengthened
by deriving predicted observables directly from the simulations.  We do not observe densities, temperatures
and pressures of clouds directly---rather, we rely on spectral line and continuum processes to 
deliver photons containing the valuable
information, from which we infer the state of the object under study.  As radiative transfer is a guiding
principle for this propagation of information, we endeavour to simulate the creation of observable photons
via radiative transfer, within the modelled physical representation of the ISM in our Galaxy.  \citet{gib06}
demonstrated the production of HISA by spiral shocks with a two-dimensional, purely atomic (cold and warm {\sciHI})
radiative transfer model.

In this paper we present results from generating synthetic {\sciHI} spectral line datacubes from a multiphase
numerical galaxy calculation.  The aforementioned SPH calculation by \citet{DobbsGloverClarkKlessen2008} is
used as a basis to create a synthetic Galactic Plane survey, fashioned after the {\sciHI} data products created for the
IGPS.  Simulating such surveys requires the ``observer" to be placed at an appropriate position inside the simulated 
Galaxy.  Such challenges are accomplished using {\Tor}, an Adaptive Mesh Refinement (AMR) code which is capable of
performing radiative transfer calculations to produce synthetic spectra \citep{Harries2000}.
Moreover, the added flexibility of 
operating on a numerical model enables us to use the same code to observe the Galaxy from multiple positions,
as well as to emulate extragalactic {\sciHI} observations \citep{Acr10}.

This paper is organised in the following way.  Section 2 reviews the method used to create our synthetic
Galactic Plane Survey, and the data cubes produced are described in section 3.  In section 4 we discuss the use of
HISA profiles as a diagnostic of early molecular cloud formation.  In section 5 we compare briefly our
synthetic data to observational {\sciHI} profiles of the Milky Way, setting the basis for a discussion of how
future refinements to our investigation, such as the incorporation of feedback mechanisms, will bring closer
agreement between the synthetic and observed {\sciHI} distributions. 

\section[]{Numerical Work: SPH Simulations and TORUS}

The method for creating synthetic {\sciHI} cubes has been outlined previously in \citet{Acr10}, hereafter called
Paper I.  We briefly summarise the main points here.
 
An SPH spiral galaxy calculation is fed into {\Tor}, wherein the
particle representation of the gas is transformed into a gridded representation, aided by AMR
to provide the best resolution in regions of highest density.  Next, radiative transfer
calculations and ray tracing algorithms produce synthetic spectra for lines of sight chosen from an
arbitrary observing position.  The intensities determined by the radiative transfer and ray tracing are
converted to a brightness temperature $T_b$, using the Rayleigh-Jeans approximation.
In Paper I we demonstrated this method by simulating 21-cm {\sciHI}
observations of nearby spiral galaxies, assigning the SPH galaxies orientations and systemic velocities
matching M31 and M33.

Moreover, the observer can be placed within the galaxy, allowing for the simulation of {\it Galactic}
{\sciHI} observations to be performed.  In such cases, some adaptations of {\Tor} algorithms must be made. 
The AMR grid is generated at a much finer resolution than that used in Paper I, in order to provide better 
resolution for structure close to the observer.  The maximum {\sciHI} mass per cell for the AMR grids used in 
this paper is $2.5\times 10^{34}$ g ($12.6$ $M_\odot$), which is a factor of 100 smaller than that used in paper 
I.  In order to cope with the increased number of cells in the AMR grid, the domain of the grid is restricted to 
exclude regions of the galaxy which are not relevant to the cube being calculated.  A different AMR grid is used 
for generating each data cube but they typically comprise $6\times10^6$ unique voxels.

Also, while the same ray-tracing method is used for both the external and internal views, its execution must be
handled differently if we are to consider multiple lines of sight.  For the external case the rays are essentially
parallel, whereas in the internal case there is an obvious angular dependence to the observations.  
Additionally, by reversing the direction of ray trace, relative to that employed in 
Paper I (i.e.~back to front rather than front to 
back), we are able to calculate the contribution of each cell of the AMR grid to either the emission or absorption in 
a given data cube pixel.  This feature is used to decompose spectral features into separate emission and absorption 
components, as part of our investigation of HISA in spiral arms.  These 
differences are critical when we extend the creation of single spectra to making fully three-dimensional datacubes
of simulated {\sciHI} emission.  The cubes are discussed in the next section.

\begin{figure}
 \includegraphics[angle=-90,width=84mm]{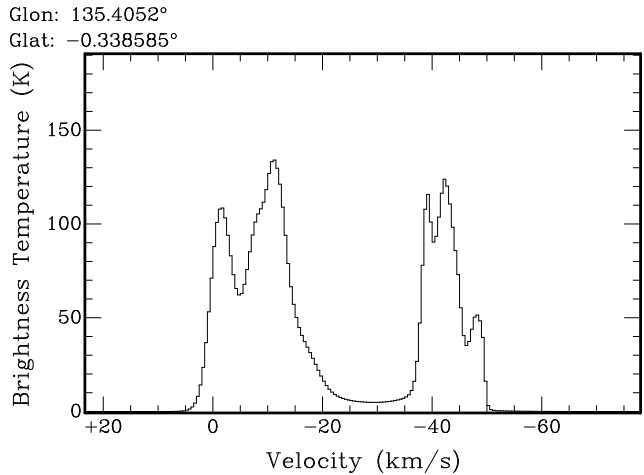}
 \caption{A synthetic {\sciHI} spectrum measured from within the model galaxy, toward a position
near Galactic coordinates $\ell = 135.5^\circ, b = -0.3^\circ$.}
 \label{figsp1}
\end{figure}

Figure \ref{figsp1} shows an {\sciHI} spectrum measured from a position within
the SPH galaxy, corresponding roughly to the Earth's position in our Galaxy, toward a direction corresponding 
to Galactic longitude $\ell \approx 135.5^\circ$ and latitude $b \approx -0.3^\circ$.  We see ``local" spiral arm gas
in a radial velocity range $-20 < v_r < 0$ km s$^{-1}$, and an outer arm beyond $v_r \approx -38$ km s$^{-1}$.  The
outer arm shows some evidence of {\sciHI} absorption, with a narrow dip in $T_b$ appearing near $v_r = -41$ km s$^{-1}$.
This HISA signature is common in synthetic spectra all across our region of interest.
Since HISA is a main focus of our investigation into molecular cloud formation, the {\Tor} code was modified so that
in addition to total brightness temperature cubes, we create separate datacubes where only positive or negative
intensity contributions ($dI$) for a given grid cell are mapped.  

For this paper, the simulated galaxy has reached a time of 250 Myr. At this age the
distribution of gas in different phases, and the fraction of molecular gas, have reached a roughly steady state.
While tracing the evolution of specific ISM structures using many timesteps from an SPH simulation is
the subject of future investigations, the present work concentrates on deriving {\sciHI} properties from a
single epoch, as is the case with most Galactic radio observations.
As mentioned previously, the models are restricted to Galactocentric radii of 5 to 10 kpc.  For this reason 
it is reasonable to aim to reproduce observations of the Outer Galaxy, toward its 
Anticenter ($\ell = 180^\circ$).  
In this paper, we survey the entire second quadrant of our SPH galaxy, covering 
longitudes between $\ell = 90^\circ$ and $\ell = 180^\circ$.


\section{Synthetic Datacubes of {\sciHI} Brightness}

\begin{figure}
 \includegraphics[width=84mm]{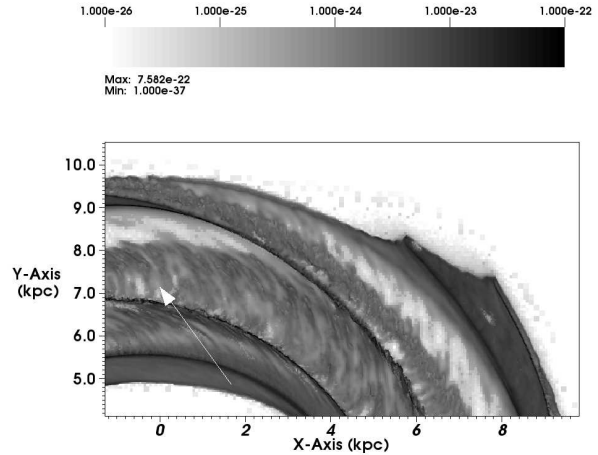}
 \caption{A midplane density slice of the simulated Galaxy gridded by {\Tor}, and showing the position of
our ``observer" for the synthetic Galactic Plane Survey.}
 \label{figobs}
\end{figure}

For the Galactic
point of view, we construct $T_b(\ell,b,v_r)$ cubes with angular coordinates corresponding to ``Galactic"
coordinates from the position of the observer.  This position is chosen to be in the midplane, $2.2 \times 10^{22}$
cm (7.13 kpc) from the 
galaxy's center, as illustrated in figure \ref{figobs}.  This choice corresponds to a position in an
interarm ``spur" of {\sciHI}, similar to the Orion Spur of our Milky Way. 

\begin{figure}
 \centering
 \includegraphics[angle=0,scale=0.7]{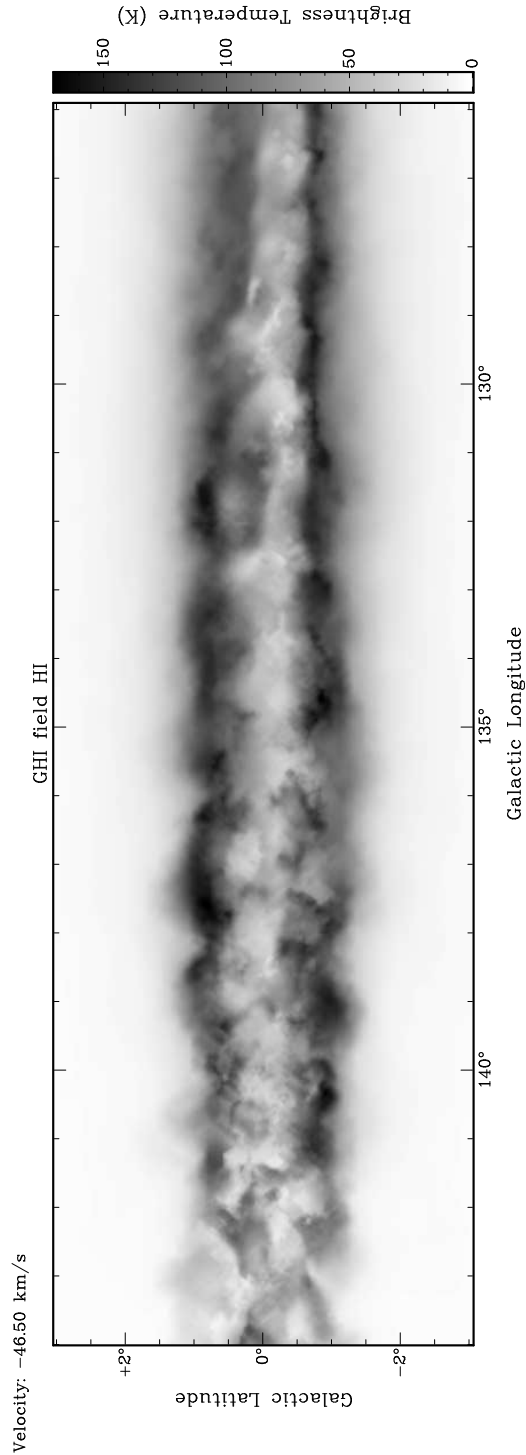}
 \caption{Synthetic Perseus Arm (SPA) gas near $v_r = -45$ km/s for the GHI region ($126^\circ \le \ell \le
144^\circ$).}
 \label{figgalxy}
\end{figure}

\subsection{Channel Maps}

A total of fifteen cubes of angular dimension $6.08^\circ \times 6.08^\circ$ are produced to cover the entire 
Second Quadrant, with a small amount of overlap between adjacent fields for mosaicking purposes.  With an image
scale of $0.5'$ per pixel, each channel map produced by {\Tor} measures $732^2$ pixels.  Our spectral range
covers radial velocities from $+30$ km s$^{-1}$ to $-120$ km s$^{-1}$ sampled at $0.5$ km s$^{-1}$, resulting
in 300 channels for each cube produced by {\Tor}.

\begin{table}
 \centering
 \begin{minipage}{240mm}
  \caption{Second Quadrant Mosaic Regions}
  \label{tblreg}
  \begin{tabular}{@{}llrrrrlrlr@{}}
  \hline
  Region Name & $\ell_{central}$ ($^\circ$)\\
  \hline
ABC & $99.0$\\
DEF & $117.0$\\
GHI & $135.0$\\
JKL & $153.0$\\
MNO & $171.0$\\
\hline
\end{tabular}
\end{minipage}
\end{table}

For ease of visualisation, the fifteen cubes are combined into five mosaics, summarised in table \ref{tblreg}.  Each
mosaic region therefore covers just over $18^\circ$ of Galactic longitude, centered on the value of $\ell_{central}$
given in the table.  For illustrative purposes we show a channel map from the central mosaic, region GHI,
in Figure \ref{figgalxy}.  This channel map shows {\sciHI} gas near $v_r = -45$ km s$^{-1}$,
as our aim is to study HISA produced in spiral arm gas with radial velocities
corresponding to Perseus Arm gas (we will refer to the simulation's ``Perseus Arm" as the ``SPA").  



\begin{figure}
 \includegraphics[angle=-90,width=86mm]{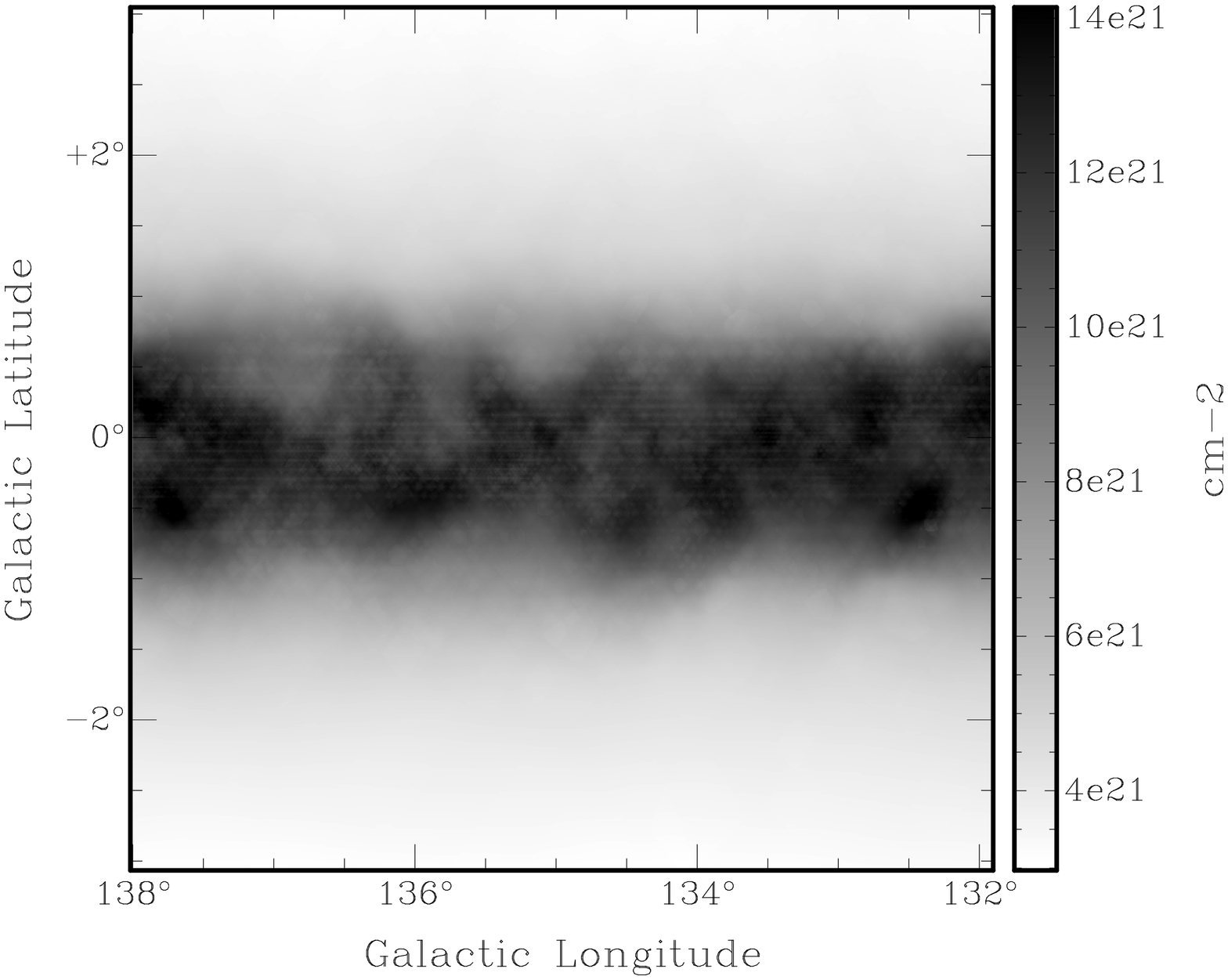}
 \includegraphics[width=83mm]{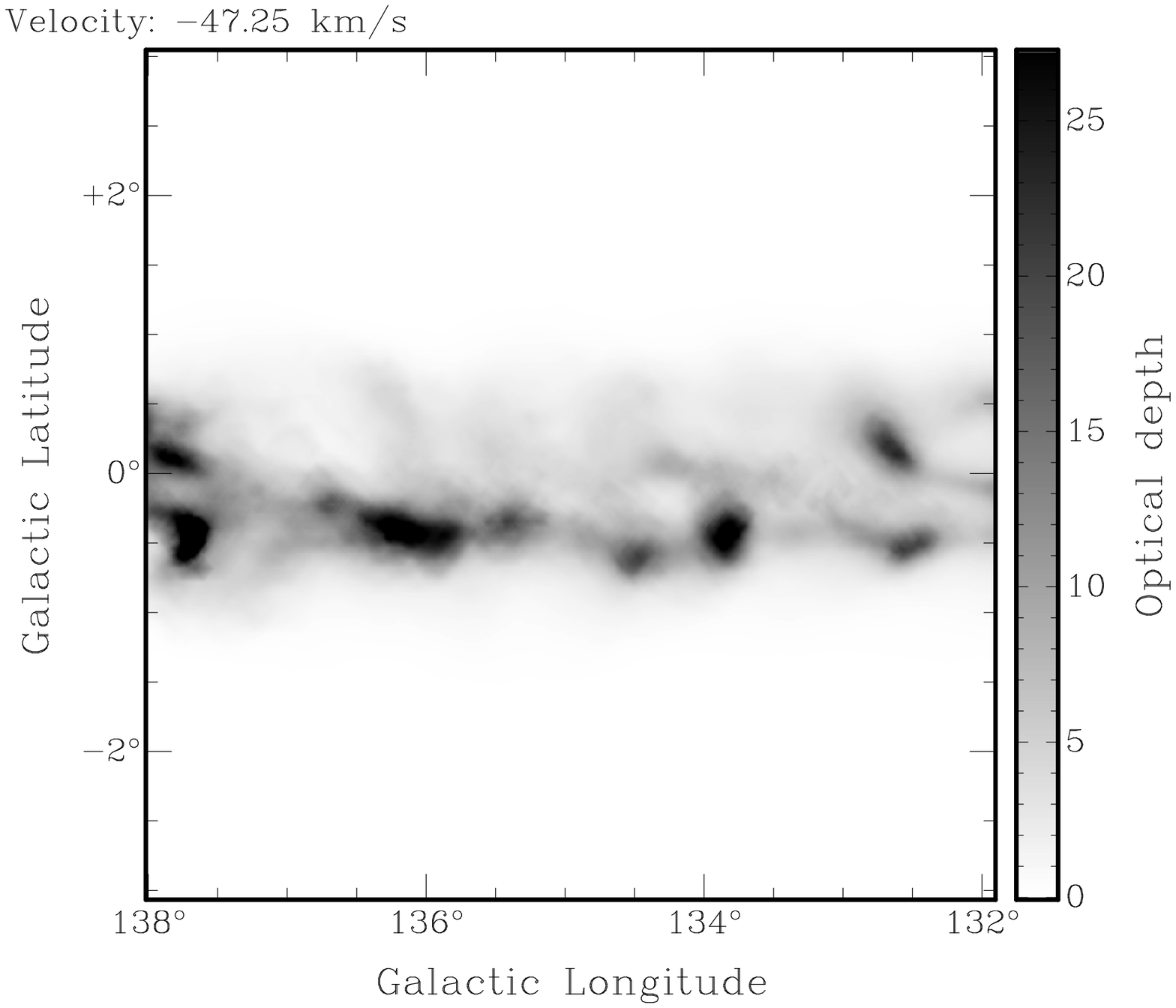}
 \caption{Column density map (top) and optical depth channel map (bottom) for the SPA cube centered on 
$\ell = 135^\circ$, $b = 0^\circ$.}
 \label{figcb8ncol}
\end{figure}

A common feature seen in our SPA channel maps is that the brightest {\sciHI} emission appears not in the central 
midplane ($b = 0^\circ$) of the spiral arm, where the densities are greatest, but in ridges at approximately 
$\pm 1^\circ$ above and below the midplane.
From the known distribution of gas column density in the original SPH simulations, the expected {\sciHI} distribution 
assuming optical thinness would not have predicted such a pattern for maximal brightness temperatures.  
Rather, this is an absorption effect arising from a highly dense and cold central ($|b| < 0.5^\circ$) midplane.  As figure
\ref{figcb8ncol} (top panel) demonstrates for the cube toward $\ell = 135^\circ$, the synthetic {\sciHI} clouds occupy 
the plane quite randomly, mostly within $|b| < 1^\circ$.  
With an increase in column density, an increased optical depth $\tau$ of {\sciHI} follows naturally.  As a result, the 
midplane of the
SPA is largely optically thick.  With {\Tor} we can construct cubes of $\tau(\ell,b,v_r)$ in addition to $T_b$,
and we show an example channel map of $\tau$ in the SPA for the same region in the lower panel of figure \ref{figcb8ncol}.
Individual {\sciHI} clouds 
exhibit optical depth values $\tau \ge 15$, and thus the {\sciHI} gas nearer the midplane absorbs much more strongly, and the
brightness temperatures are diminished by the combination of high opacity and colder temperatures.  The optical depths
above $|b| = 0.5^\circ$ are considerably lower, and so we observe emission rather than absorption.



Another observed (and related) consequence of the overdense, high-$\tau$ SPA plane is that the amount of absorption 
seen in the SPA greatly exceeds the HISA distribution found in the Perseus Arm of the Milky Way.  While
this level of self-absorption is greater than what is expected from Galactic studies \citep{Gibsonetal2005,gold07}, 
the exaggerated contrast between
emission and absorption in the SPH galaxy allows us to investigate the HISA phenomenon in a more ``controlled" interstellar
environment.  
As mentioned above, we make separate cubes which track pixels where brightness temperature increases or decreases
along the line of sight, as a means of tracing regions where self-absorption occurs.

\begin{figure*}
 \includegraphics[angle=-90,width=84mm]{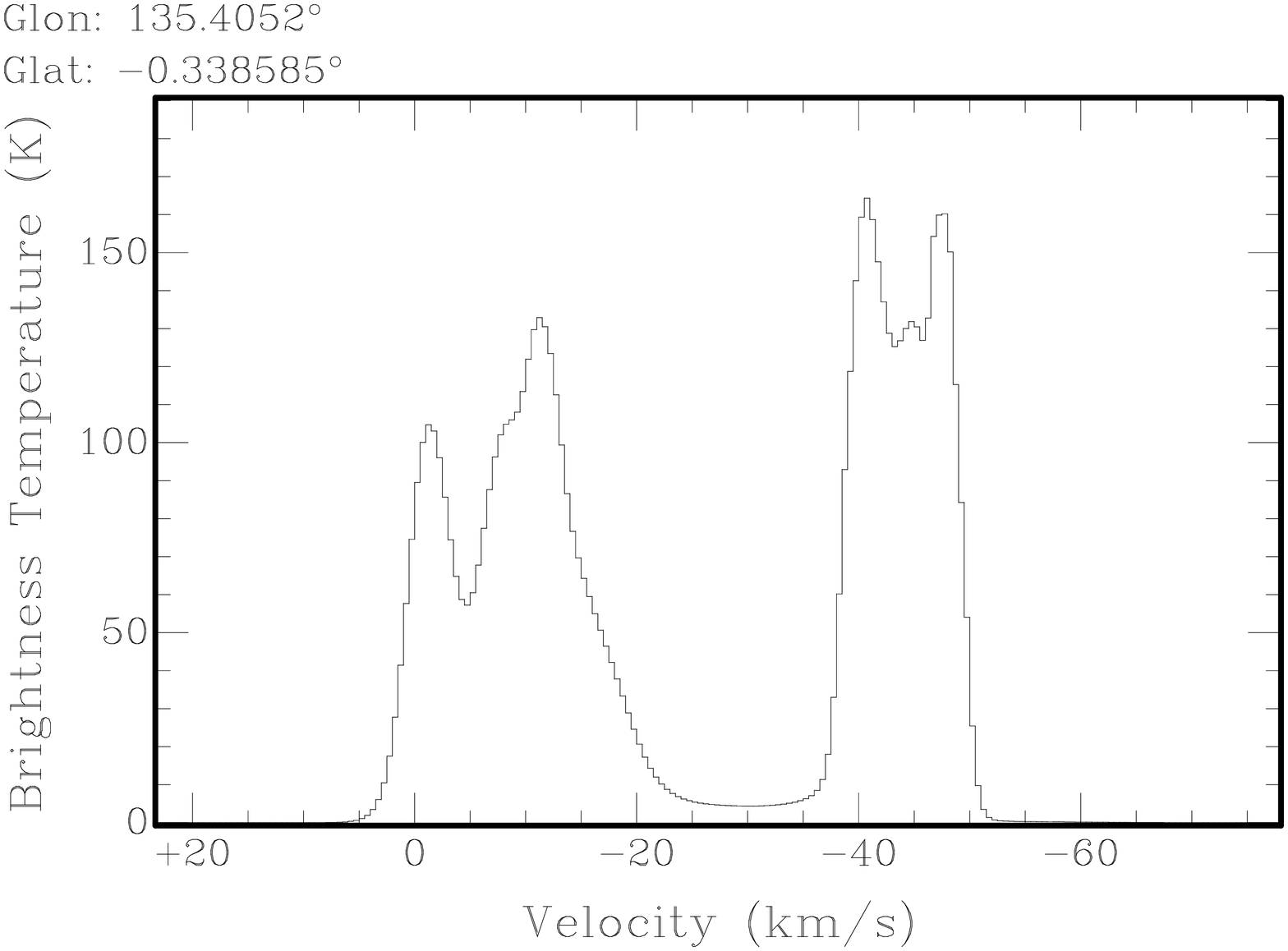}
 \includegraphics[angle=-90,width=84mm]{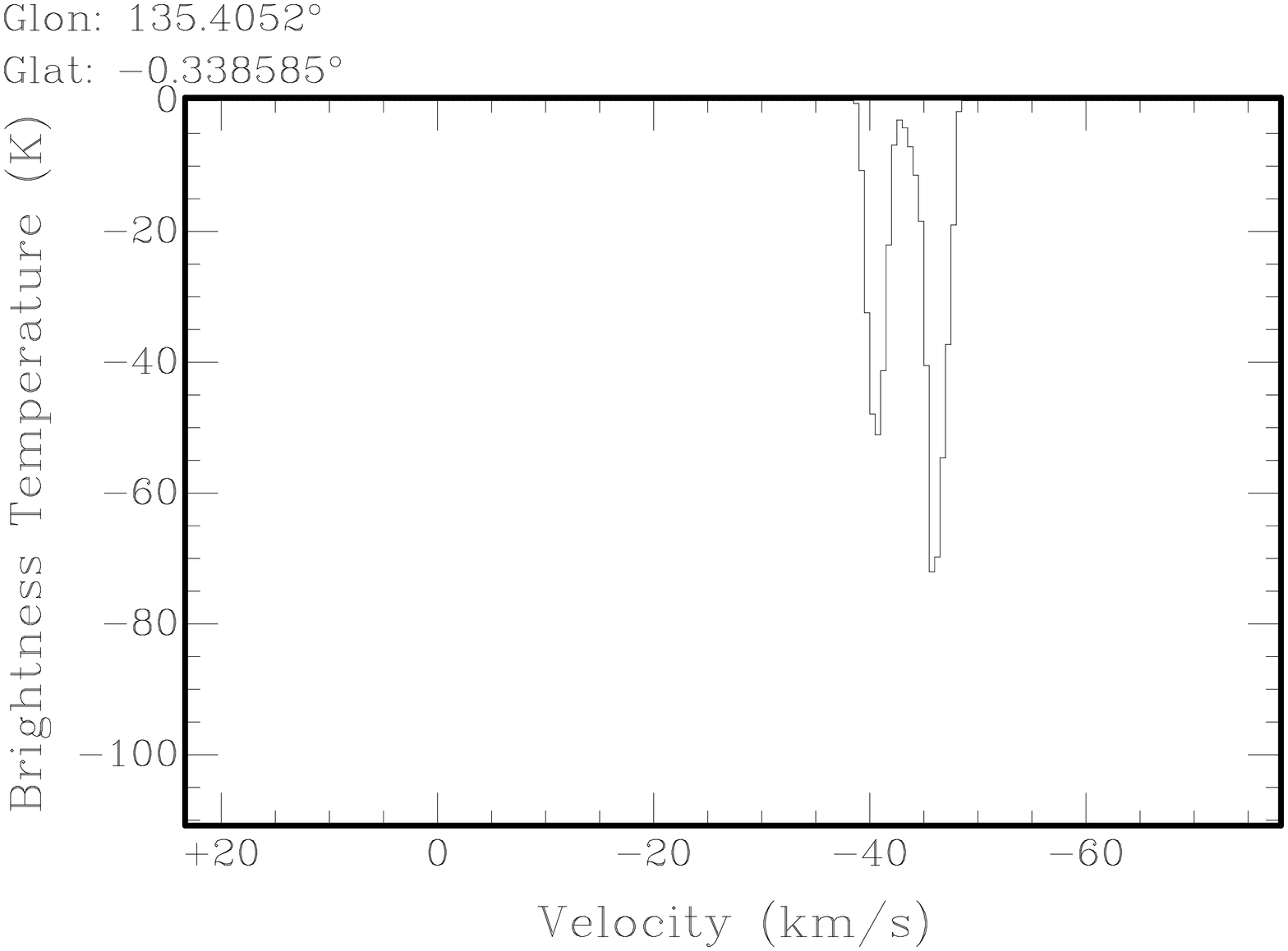}
 \includegraphics[angle=-90,width=84mm]{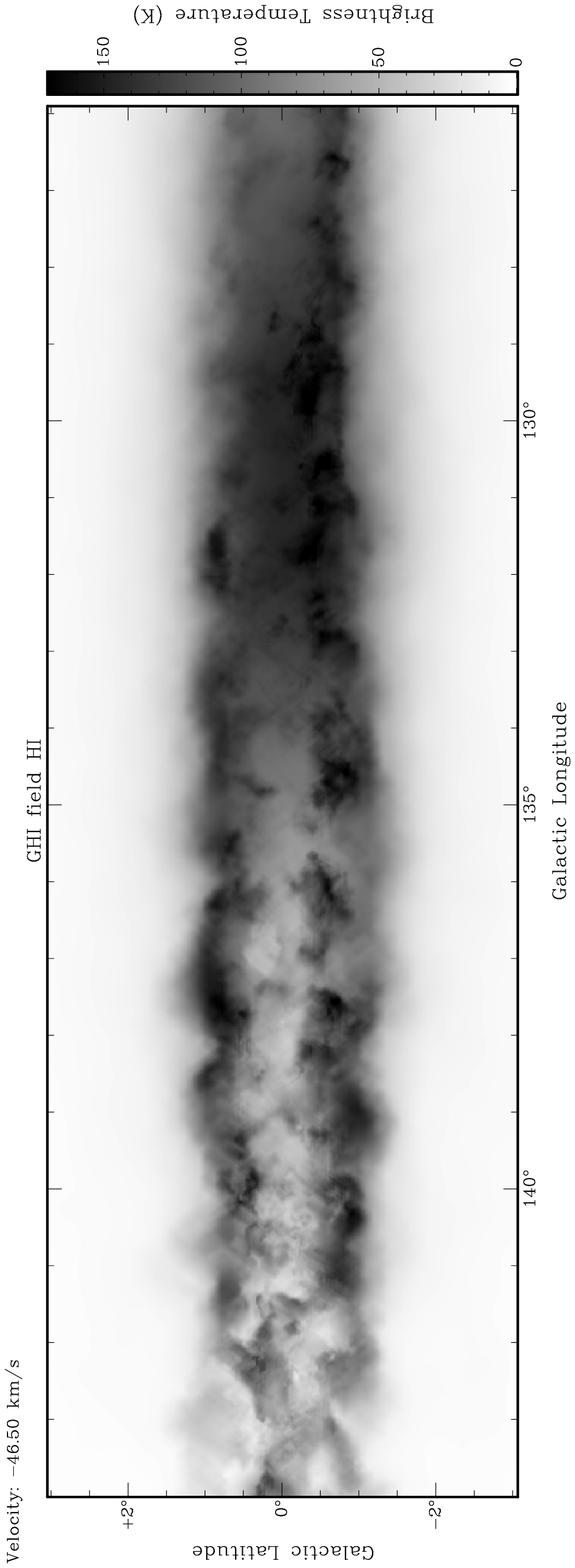}
 \includegraphics[angle=-90,width=84mm]{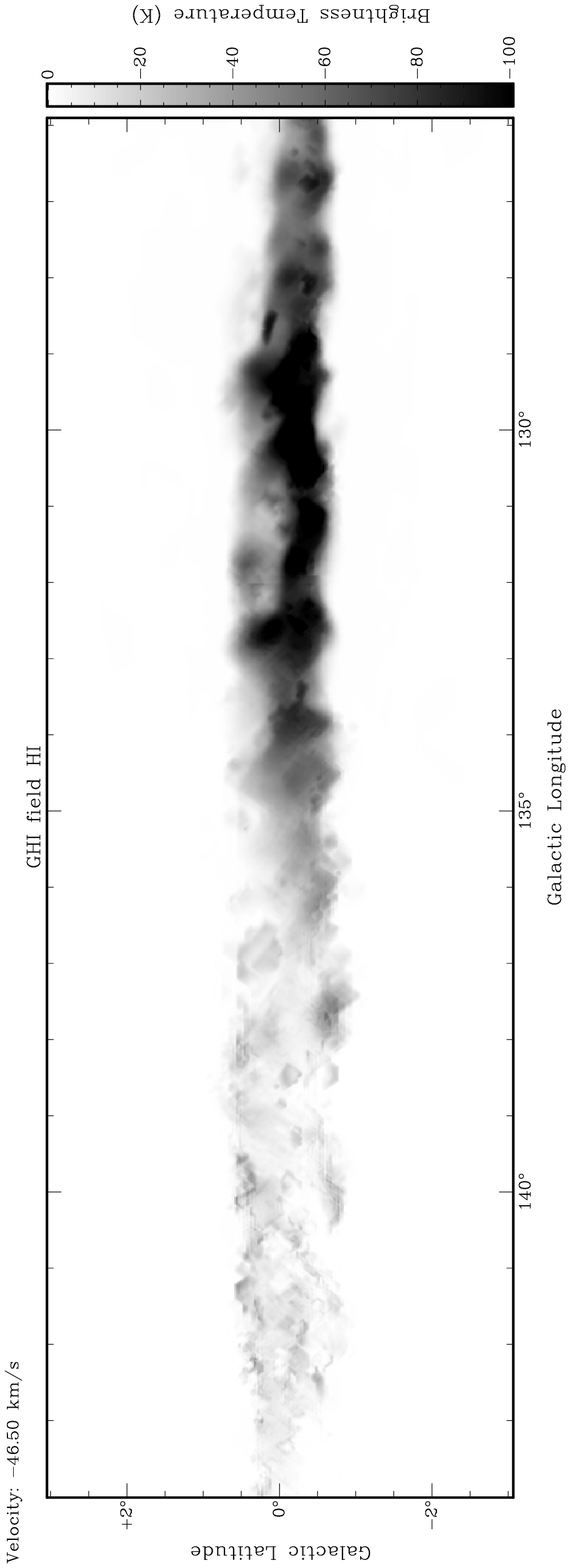}
 \caption{Synthetic {\sciHI} emission-only (left) and absorption-only (right) spectra and channel maps.  The spectra 
are for the same line of sight as in figure \ref{figsp1}, but with the positive
and negative $T_b$ contributions separated.  The channel maps show the same, for all of region GHI (see figure 
\ref{figgalxy}).}
 \label{figsp1np}
\end{figure*}

Figure \ref{figsp1np} shows positive-only and negative-only
spectra for the line of sight total-$T_b$ spectrum shown in figure \ref{figsp1}, and channel maps for region GHI at
the same radial velocity as shown in figure \ref{figgalxy}.  The negative-only spectra and channel maps are very
useful in isolating HISA features in our synthetic survey; the majority of negative-$T_b$
pixels lie in the plane at radial velocities corresponding to the SPA.  We will return to our analysis of HISA
features in the next section.

\subsection{Position-Velocity Views}

\begin{figure*}
 \centering
 \includegraphics[angle=-90,scale=0.65]{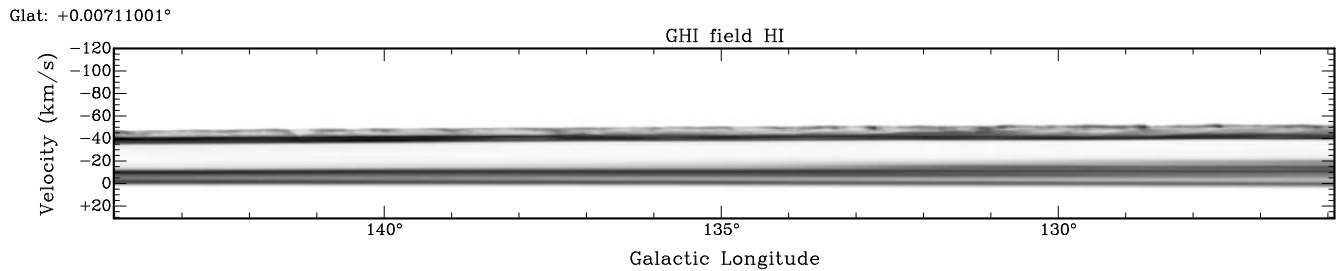}
 \caption{Longitude-velocity diagram of simulated {\sciHI} emission near $\ell=135^\circ, b=0^\circ$ for the 
Galactic synthetic
observation.  Local and Perseus Arm gas are well separated.}
 \label{figgalxz}
\end{figure*}

In figure \ref{figgalxz} we show the Galactic view toward $\ell = 135^\circ$ again (region GHI), with an $\ell-v_r$ cut 
along $b=0^\circ$.  The main feature to be elucidated is the separation of local ($v_r \approx 0$) {\sciHI} from gas 
at synthetic Perseus Arm (SPA) velocities, near $-35$ to $-55$ km/s.  The local gas has little discernable structure in our 
synthetic cube, as its proximity to the observer position means the gas has a very large angular extent.  As is the case
with our Galaxy, the mean radial velocity of the SPA becomes less negative toward higher longitudes, 
and blends with the local gas toward the Anticenter, a projection effect based on Galactic rotation. 
Toward lower longitudes, the SPA spans a larger range of radial velocities.


Beyond the SPA, there is no outer spiral arm across the Second Quadrant, until the appearance of
terminating spiral arms from spiral structure originating toward the Inner Galaxy.  Across the Second Quadrant
there are discrete clouds, containing high fractions of molecular gas in many instances,
which exit the outer edge of the SPA, but these clouds mostly share radial velocities with {\sciHI} gas within the
SPA.  

\begin{figure}
 \includegraphics[angle=0,width=84mm]{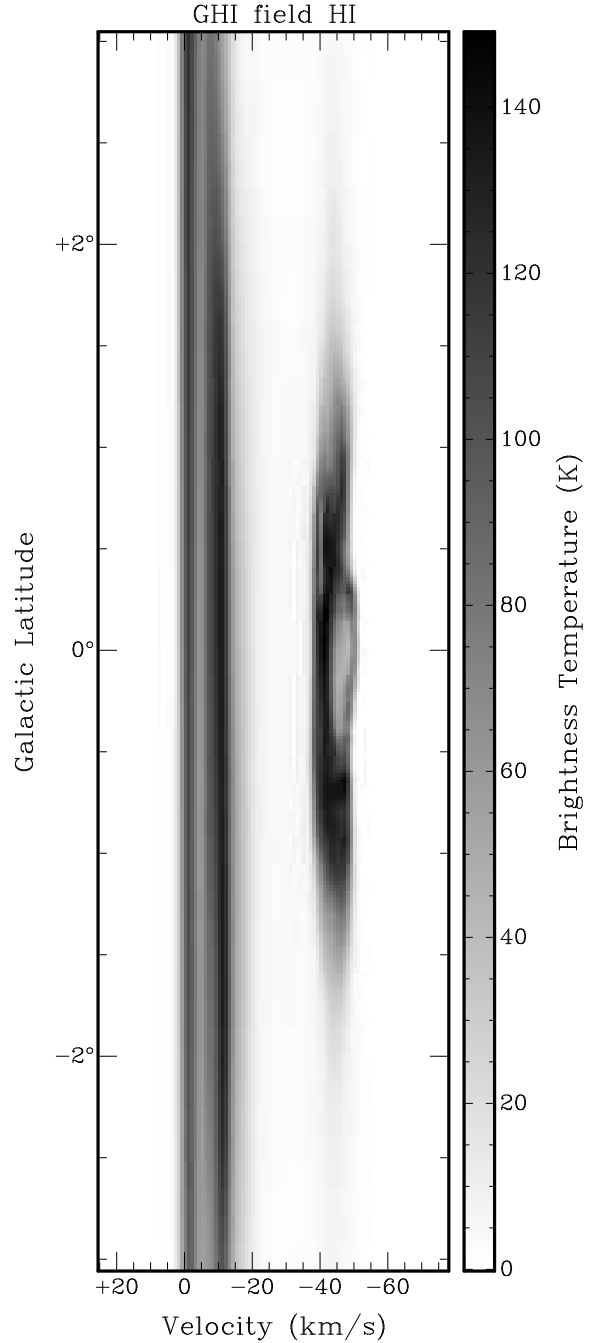}
 \caption{Latitude-velocity slice of synthetic {\sciHI} toward a Galactic longitude of $\ell = 135^\circ$.  
The SPA {\sciHI} emission is mostly confined to within 1 degree of the midplane.}
 \label{figBV}
\end{figure}

As we move away from the midplane in the longitude-velocity projection, the SPA shows bright {\sciHI}
emission out to $|b| \approx 1.3^\circ$, after which its intensity drops sharply as one moves further from
the midplane. This effect is shown toward $\ell = 135^\circ$ in figure \ref{figBV}.
With a distance to the SPA of 2 kpc (at its closest), this vertical distribution suggests that most of the 
{\sciHI} is confined to within $\sim 100$ pc of the midplane.

\section{HISA and spiral shocks}

In our Galaxy, HISA features tend to outline individual clouds or complexes of cold {\sciHI}  
\citep{Gibsonetal2005,kav05,mcc06}.  By contrast, our
synthetic data show large bands of HISA that subtend much larger angular extents.  While unrealistic, this large
abundance of HISA gives us many sightlines to test its origin.

\begin{figure}
 \includegraphics[angle=0,width=84mm]{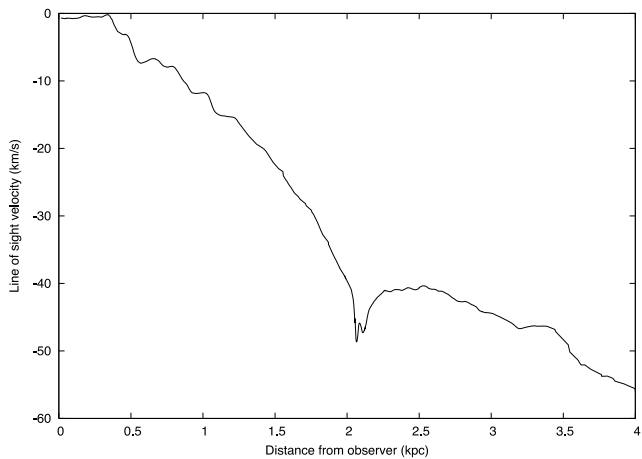}
 \caption{Radial velocity of the gas plotted against distance, for $\ell \approx 130.5^\circ$, $b \approx -0.38^\circ$
in the simulated galaxy.}
 \label{figvr}
\end{figure}

The scenario described by \citet{Roberts1972} for shock-induced compression of {\sciHI} (as evidenced via HISA)
can be tested directly with our simulations.  Figure
\ref{figvr} shows a plot of the radial velocity for gas along a line of sight toward a position in region GHI, with
$\ell \approx 130.5^\circ$ and $b \approx -0.38^\circ$.  Between the observer and a distance of about 2 kpc, the
radial velocity of the gas decreases steadily as a result of the gas moving with the rotation of the galaxy.  Then,
at a distance of 2 kpc, the radial velocity undergoes a sharp decrease from about $-40$ km s$^{-1}$ down to a local
minimum of approximately $-50$ km s$^{-1}$.  The radial velocity remains in this range as we continue to move away
from the observer, until about 2.2 kpc, where it slowly rises to $-40$ km s$^{-1}$ and levels off for about half a
kiloparsec, before again decreasing very gradually beyond $2.7$ kpc.

\begin{figure}
 \includegraphics[angle=-90,width=84mm]{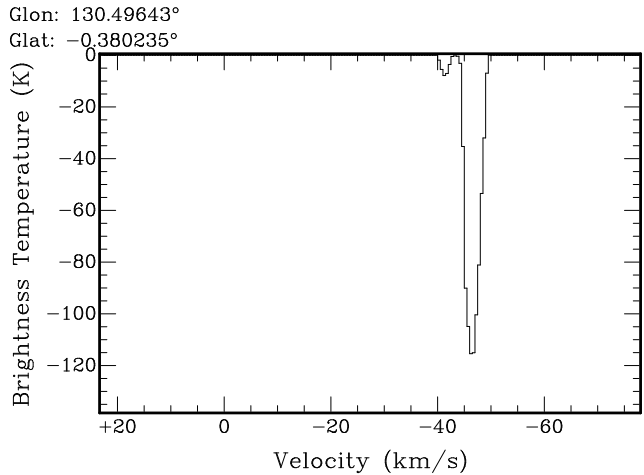}
 \caption{The HISA profile (ie.~negative-$T_b$-only spectrum) toward the position described in figure \ref{figvr}.  The
radial velocities where
HISA occurs are between $-40$ and $-50$ km s$^{-1}$, exactly where it would be expected from figure \ref{figvr}.}
 \label{figHISAvr}
\end{figure}

The dip near 2 kpc corresponds to the near edge of the SPA, and we interpret this sharp decline in radial
velocity as the result of the {\sciHI} gas experiencing a shock in the spiral arm, where it loses much of its
line of sight velocity as it begins to travel along the spiral arm edge.  The gas cools during this process,
and so the hydrogen at 2 kpc will absorb the {\sciHI} emission from the gas behind it with the same radial velocity,
resulting in strong HISA present in the spectrum between $-40$ and $-50$ km s$^{-1}$.  Indeed we show in figure
\ref{figHISAvr} that this is the precise radial velocity range where HISA is prevalent at this same position in the
datacube.  This confirmation of the \citet{Roberts1972} scenario for {\sciHI} shocks, in a region where molecular clouds
are also formed in the model, shows the value of HISA as a tracer of the atomic-molecular interface of the ISM.

\begin{figure*}
 \includegraphics[angle=0,width=84mm]{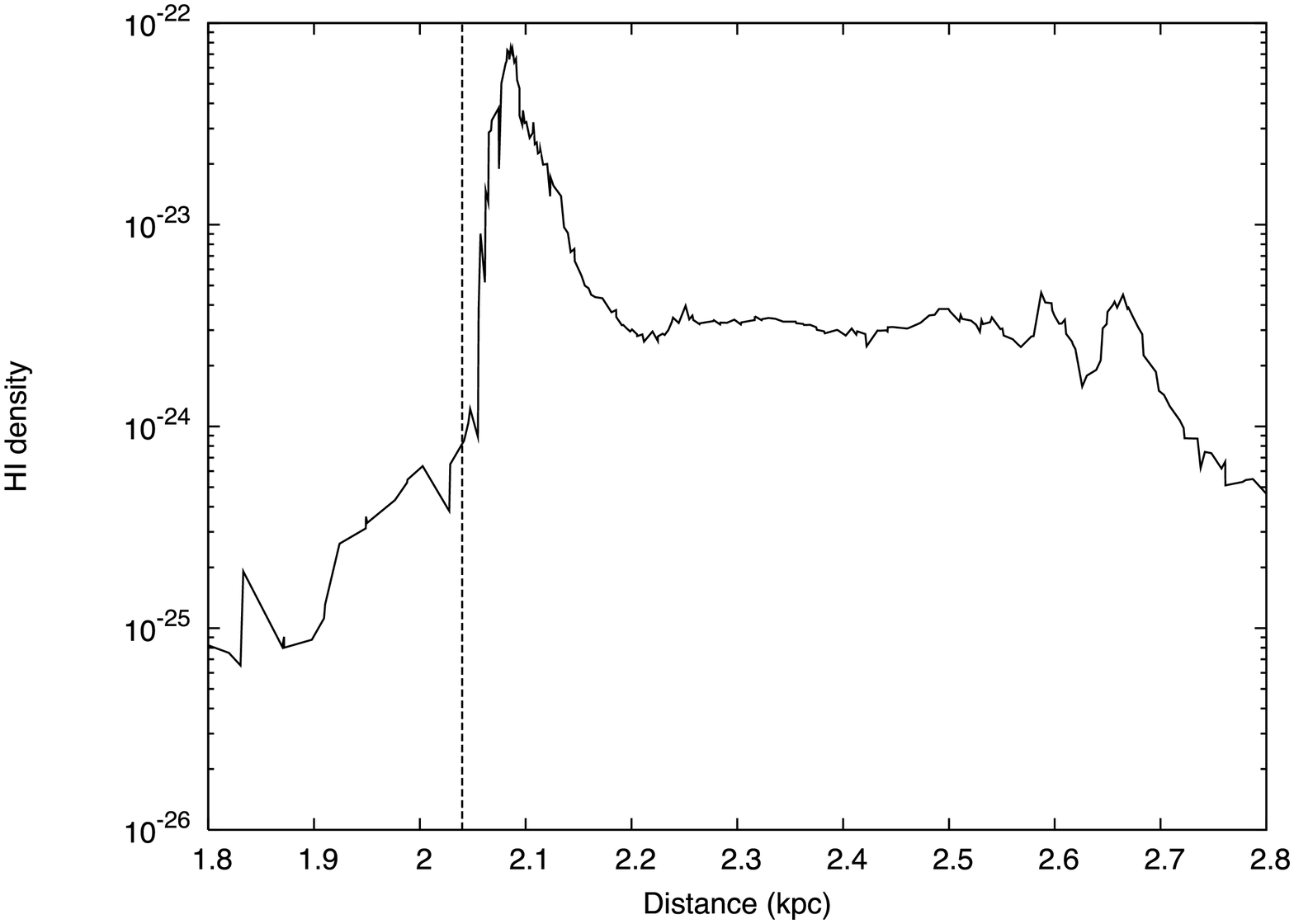}
 \includegraphics[angle=0,width=84mm]{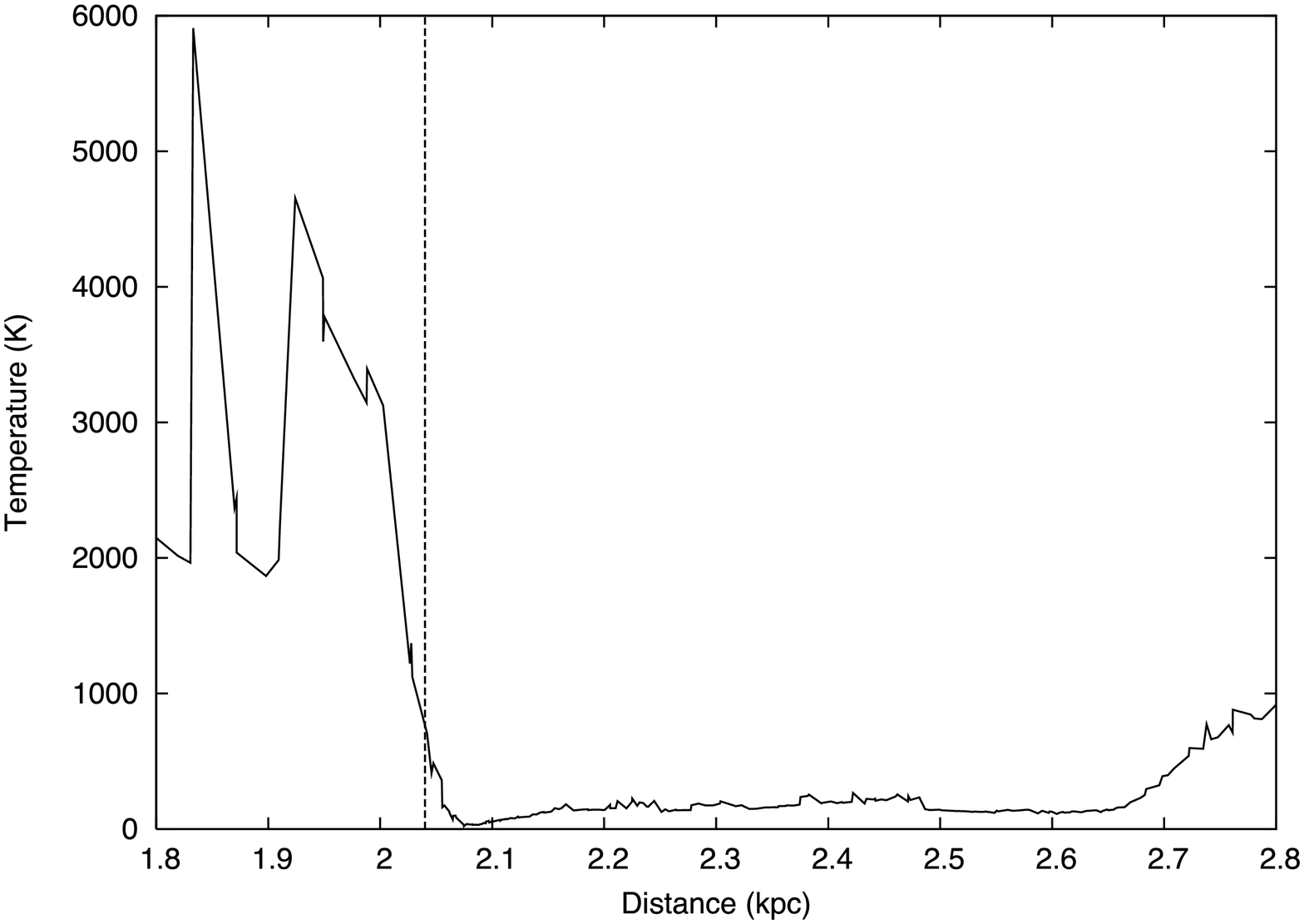}
 \includegraphics[angle=0,width=84mm]{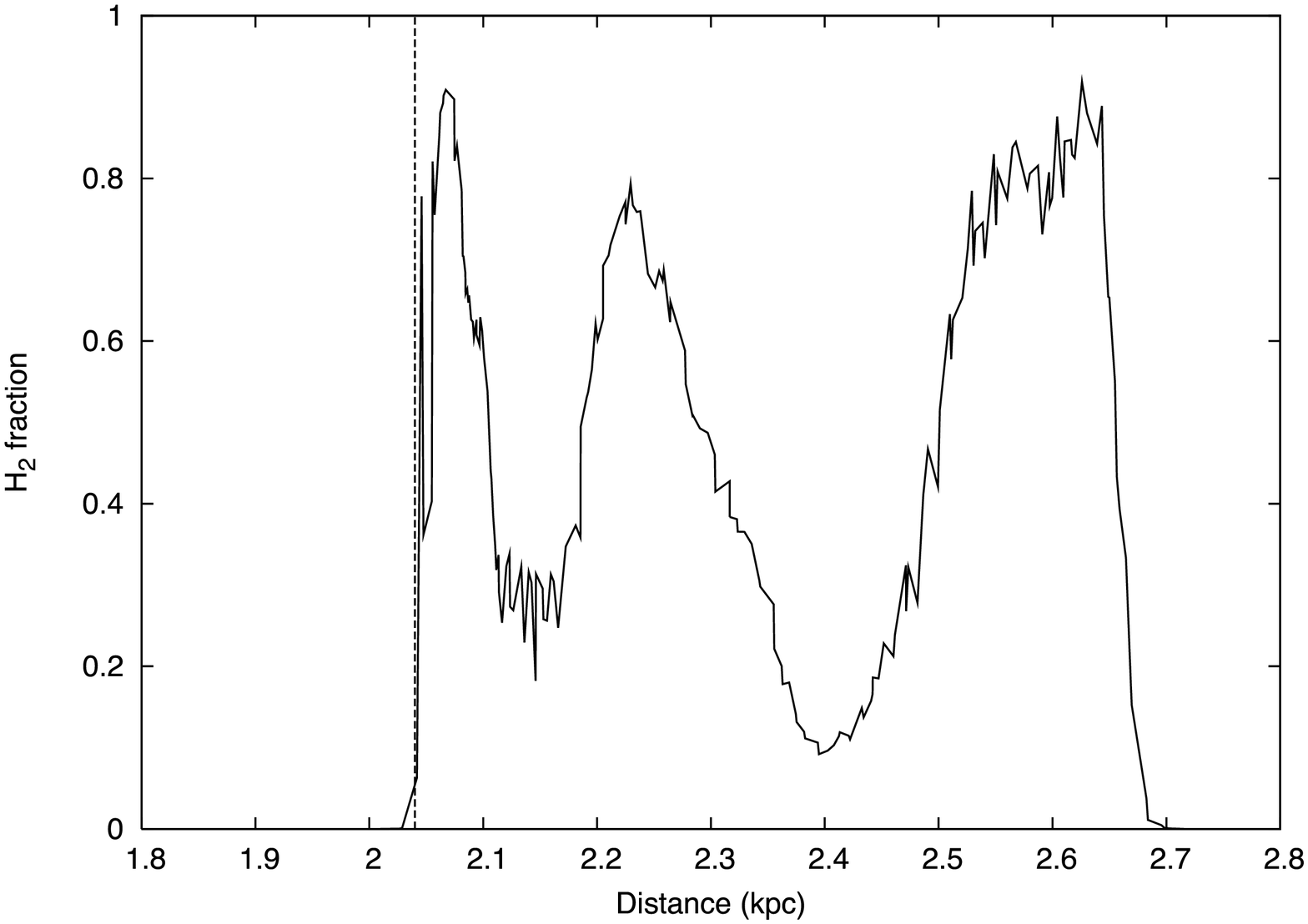}
 \includegraphics[angle=0,width=84mm]{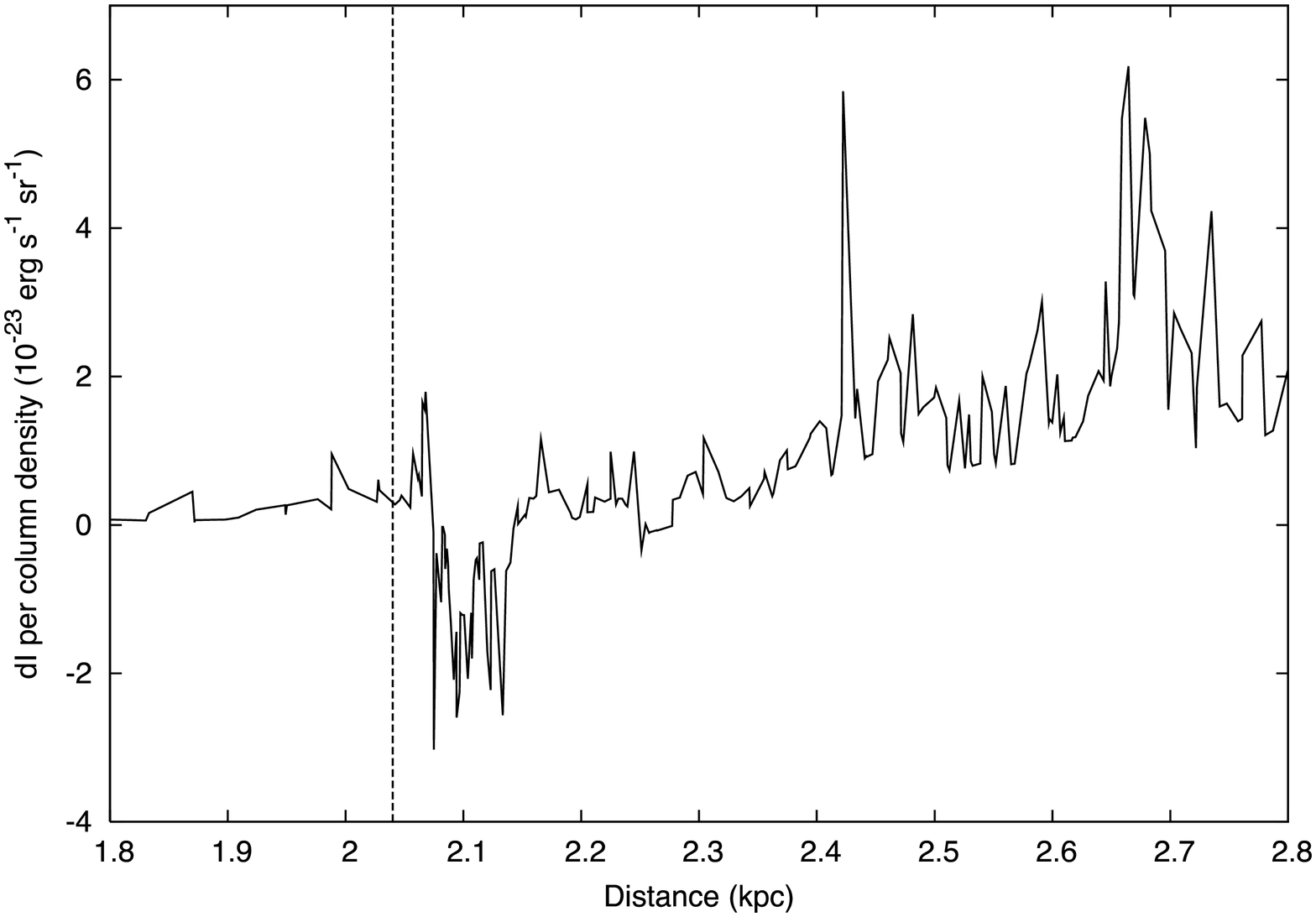}
 \caption{A close-up of the shock region seen in figure \ref{figvr}.  The shock front position is approximated by the vertical
dashed line.  As {\sciHI} density increases (top left) 
and temperature decreases (top right) , H$_2$ is formed in the shock region (bottom left), and background {\sciHI} emission 
is absorbed (negative $dI$ values in bottom right panel).} 
 \label{fighisa}
\end{figure*}

In figure \ref{fighisa} we zoom in to distances corresponding to the SPA, for the same line of sight as figure \ref{figvr}.
While this sightline shows appreciable HISA, there is nothing extraordinary about it compared to other regions across
the SPA.  It is a fair representation of the type of {\sciHI} profiles observed across the midplane of our synthetic survey.
As shown in the top left panel, the density of neutral hydrogen rises by two orders of magnitude between 2.0 and 2.1 kpc from 
the observer, as the spiral arm boundary is crossed.  While a single-fluid ($\gamma = \frac{5}{3}$) adiabatic shock would
result in at most a factor of four density enhancement, such conditions would not constrain a strongly cooling shock, which
describes the present scenario.  Indeed, a corresponding drop in gas temperature is shown in the top right panel,
as the gas cools strongly from values of $\sim 3000$ K down to less than 100 K.  In the lower left figure, we plot the 
H$_2$ fraction, by mass, of the total gas.  The first peak, between 2.0 and 2.1 kpc, is interpreted as newly formed H$_2$ 
that has been formed in the vicinity of the shock at that site in the SPA.  The other two high-H$_2$ areas beyond the shock,
near 2.25 and 2.6 kpc, would be (giant) molecular clouds that were formed upstream by the same shock mechanism, at higher longitude, 
and which are moving through the SPA.  In the absence of feedback in our simulation, these clouds may be larger than expected,
and longer-lived.  The outer edge of the SPA is less easy to define than the inner edge, given that there is no sharp outer
boundary, but across much of the SPA, the density begins to decrease toward interarm values approximately 0.7 to 0.8 kpc behind
the inner boundary.

The effect of these density and temperature variations on the observed {\sciHI} profile can be informed by the lower right
panel of figure \ref{fighisa}, which shows the intensity contribution $dI$ per unit of column density as a function of 
distance.  HISA (negative $dI$) appears to be produced behind the shock, once the temperature has dropped to below
100 K, and where the {\sciHI} density is above $10^{-23}$ g cm$^{-3}$.  At the front of the shock there is a narrow region
where emission is produced rather than absorption, because the gas is still quite warm, yet the {\sciHI} density is already
increasing rapidly.  Such conditions are favourable for the onset of H$_2$ production, which depends strongly on density as
well as weakly on temperature \citep{Ber04}.  Still, this shows that within the region where the gas velocity is strongly
influenced by the spiral arm shock, both emission and absorption from {\sciHI} atoms are possible.  Yet we can constrain
the occurrence of HISA to cold, dense regions which also contain molecular gas.
The temperature appears to remain at values around 100 K out to a distance of $\sim 2.65$ kpc, after which it begins to rise
and the molecular fraction drops sharply.  Thus the cold gas is mostly confined to the spiral arms, and the gas is warmed
in the interarm regions.  \citet{DobbsGloverClarkKlessen2008} found that roughly 62 per cent of the {\sciHI} was in the cold
phase, where $T \le 150$ K was their criterion for cold gas.

The shock region itself is much smaller than the $\sim 100$ pc region behind it where the physical state of the ISM
undergoes these interesting changes.  In the original SPH simulations, the smoothing length decreases greatly in high-density
regions, dropping to values of about 5 to 6 pc to resolve regions where the density approaches $10^{-22}$ g cm$^{-3}$ 
\citep{dobbs08}, which is approximately the peak density achieved for the shocked {\sciHI} shown in figure \ref{fighisa}.
SPH simulations typically require several smoothing lengths to capture a shock, and the shock width in any numerical model will be 
considerably greater than any realistic shock width \citep{Price08}.

\section{Discussion}

\subsection{Comparison with Observational Data}

The synthetic datacubes we have produced correspond to the Second Galactic Quadrant, for which the best 
comparison set is the Canadian Galactic Plane
Survey, part of the International Galactic Plane Survey \citep{tayetal03}.  We compare below the general 
characteristics and features observed in both the synthetic and real data.  By the very fact that our SPH
galaxy is {\it not} the Milky Way, no strong agreement is expected {\it a priori}.  The main purpose of the
synthetic models is to investigate whether the geometry, kinematics, chemistry and thermodynamics of a spiral shock can 
produce HISA, as has been observed in the Perseus Arm of our Galaxy.
In comparing the two datasets, we can discern what characteristics this relatively simple galaxy model
has in common with Galactic {\sciHI} observations.

\begin{figure}
 \centering
 \includegraphics[angle=0,scale=0.7]{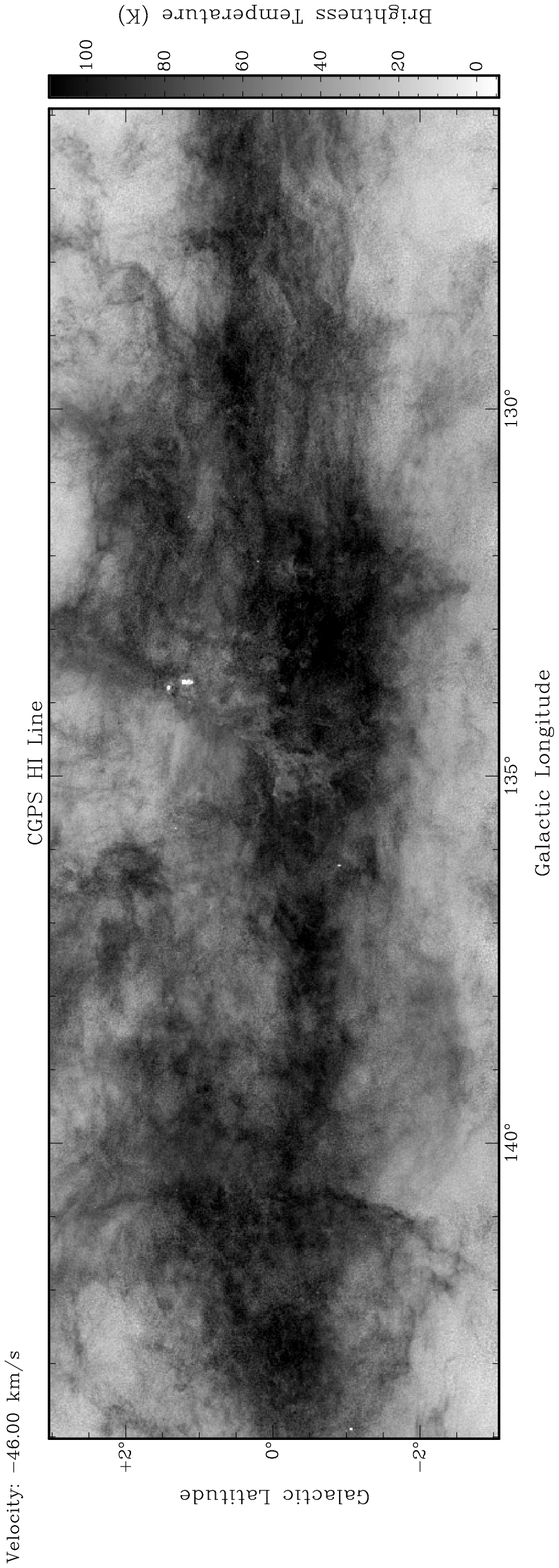}
 \caption{CGPS Channel map of Perseus Arm gas centered on $\ell = 135^\circ$, $b = 0^\circ$.}
 \label{figCgGHI}
\end{figure}


In figure \ref{figCgGHI} we show the Galactic analogue of the synthetic {\sciHI} channel map for region GHI that was shown in
figure \ref{figgalxy}.
The main difference observed across the Second Quadrant is the apparent confinement of {\sciHI} gas to the plane in our
synthetic data cubes, compared to the widespread and more diffuse distribution of atomic gas seen in the CGPS data.
This is not surprising, given the lack of feedback mechanisms in the simulations that would serve to stir up the interstellar
medium more vigourously---we will discuss this point further below.  For now, we consider the consequences of this
scale-height mismatch.  For one, with a greater column density of {\sciHI} toward the overdense midplane, the brightness
temperatures for the synthetic line profiles are larger than what is observed in the Milky Way.  Within the CGPS data 
brightness temperatures above 150 K are not common, yet in the synthetic cubes we encounter many regions with $T_b > 200$ K.


\begin{figure*}
 \centering
 \includegraphics[angle=-90,scale=0.64]{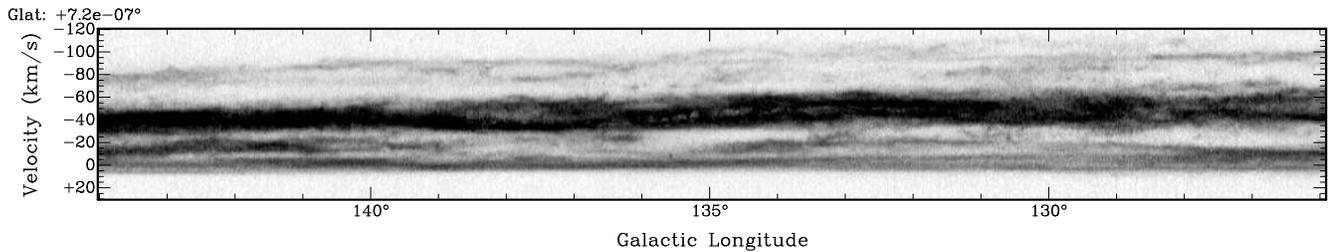}
 \caption{Longitude-velocity diagram of Milky Way {\sciHI} emission near $\ell=135^\circ, b=0^\circ$ as observed 
with CGPS data.}
 \label{figCgGHILV}
\end{figure*}

Unlike the scale height differences observed in the channel maps, the spiral structure in the SPH galaxy and the Milky Way
show good agreement, in terms of the Perseus Arm, the main feature in the Second Quadrant of the Outer Galaxy.  Figure
\ref{figCgGHILV} shows the Milky Way's mid-Second Quadrant midplane longitude-velocity distribution, which 
portrays a similar separation, on the order of $\Delta v_r \sim 40$ km s$^{-1}$, between
local and Perseus Arm gas to that observed in figure \ref{figgalxz}.  The presence of gas beyond the Perseus Arm 
is a main difference exhibited by the real data, as well as distinct {\sciHI} features in the interarm gas.  
Additionally, the overall velocity width of the Galaxy's spiral arms is in general
greater than that of the simulated galaxy. In the SPH galaxy, particles were given circular velocities with a velocity 
dispersion assigned to be $\sigma = 6 $ km s$^{-1}$,
while values of $\sigma \sim 8$ km s$^{-1}$ or greater typically describe the {\sciHI} of our Outer Galaxy \citep{Saha09}.
Many factors contribute to such a dispersion value, including feedback mechanisms and a warped disk; \citet{DobbsPrice2008} 
also found that magnetic fields broaden spiral arms in SPH galaxy simulations.




\subsection{The Role of Feedback}

The interstellar medium of our Galaxy is constantly disrupted, by both external and internal agents.  The inflow
of gas from satellite galaxies and high-velocity clouds provides fuel for future generations of molecular clouds
and stellar systems.  Numerous feedback processes from stellar evolution also add energy and momentum to the ISM
from within: outflows, supernova shocks and winds from individual stars or stellar associations.  With these 
agents absent from the simulated galaxy, our synthetic observations correspond to a rather quiescent version of
a Milky Way-type galaxy.  The consequences are abundantly clear in our synthetic datacubes.  {\sciHI} clouds collect
toward the midplane and cool, with nothing to destroy them, resulting in larger than expected column and number
densities.  Thus the lower scale height seen in our simulated galaxy is well explained.  If our synthetic data
had shown agreement with the {\sciHI} scale height suggested by the CGPS data, then the importance of feedback in
contributing to the overall structure of the ISM would be severely diminished.  

Much of the {\sciHI} distribution in our Galaxy can be described with words such as filaments, shells, and bubbles.
Interstellar bubbles are created by outward forces originating from a central nidus.  Shells and filaments 
form as a result of bubbles encountering each other, or other ISM structures, and their evolution is also aided by radiative
and magnetic field effects.  Without the injection of
supernova-like events or winds from OB associations, the simulated galaxy cannot be expected to recreate the 
filamentary structure observed with {\sciHI} observations of the Milky Way.
Hence feedback is a possible solution to both the overconfinement of {\sciHI} in the midplane, as well as
the non-filamentary nature of the {\sciHI} channel maps.
Feedback should also increase the velocity dispersion of the simulated galaxy's spiral arms, since the mixing of
the {\sciHI} will certainly result in gas with a wider range of radial velocities.

\subsection{Future Work}

The simulation used in this paper provides a basis with which to compare future simulations which include more 
physical processes.  As highlighted above, feedback appears to be instrumental in reproducing the correct structure 
(holes, filaments) and vertical distribution of the ISM.  
Moreover, magnetic fields are an important source of pressure, supporting the gas in the vertical direction 
\citep{cox05}.  In addition, magnetic fields are likely to reduce thermal instability in the gas \citep{fie65,hen00,sto09},
which could reduce the amount of HISA seen in our simulated observations.
On the other hand, self gravity will promote the formation of dense {\sciHI} or molecular structures in the midplane, 
particularly at higher surface densities.   Our synthetic observations of the {\sciHI} distribution and the
resulting HISA properties in these
future simulations will demonstrate the cumulative effect of the added physics.

As the molecular fraction of the SPH galaxy evolves, formation of H$_2$ and CO take place using prescribed
rate equations \citep{DobbsGloverClarkKlessen2008}.  The simplicity of the
two-level hyperfine 21-cm transition makes the production of the synthetic {\sciHI} spectra rather
straightforward, compared to the detailed balance calculations required for other ISM tracers.  Still,
in principle the {\Tor} code can be adapted to produce synthetic observations of any species present in the
original simulation.  \citet{run10} have achieved synthetic molecular line observations for several species,
including CO, for far-field view simulated observations of molecular clouds with {\Tor}.  We will integrate
their work to produce a synthetic CO galactic plane survey to complement our {\sciHI} data.
Besides spectral lines, we will also aim to produce realistic maps of continuum emission from dust.

The use of numerical simulations also offers the unique opportunity to trace individual features in the time
domain.  The simulation we used here from \citet{DobbsGloverClarkKlessen2008} and others like it show the
evolution of atomic hydrogen clouds as they encounter shocks in spiral arms, form molecular clouds, and eventually
disperse as they exit the spiral arm on the other side.  Radio observations can only piece together this scenario
by observing multiple clouds at different stages of this evolution.  In a future study we will recreate our
synthetic Galactic Plane survey for several timesteps of a numerical solution, and track the evolution of the
{\sciHI} profiles of gas clouds which pass through spiral shocks.

In all of the above studies, we can also extend our synthetic Galactic Plane survey to other parts of the
simulated Galaxy.  From our chosen observer position, only lines of sight within $|\ell| < 46^\circ$ are affected
by the absence in the model domain of the inner 5 kpc of the galaxy.  Thus we can extend our investigation to include 
{\sciHI} observations from other components of the IGPS for comparison.

\section{Summary}

We have demonstrated our ability to conduct a synthetic Galactic Plane {\sciHI} survey.  With the aid of the versatile
{\Tor} code, we converted an SPH simulation of a spiral galaxy section into a set of {\sciHI} datacubes predicting the
brightness temperature distribution from an observer's position within that galaxy.  With these cubes we can trace global
features such as HISA and spiral arm structure, and in comparison with Galactic Plane Survey cubes of our Galaxy, we find
common features, as well as clear and explicable differences in the {\sciHI} distribution.  The main differences seen, which 
include a lack of filamentary structure, a lower {\sciHI} disc scale height, and spiral arms with a lower velocity dispersion,
may be well explained by the absence of feedback processes that would conspire to mix the SPH galaxy's ISM more vigourously.
We have shown that HISA is produced
naturally by the influence of a spiral arm in the outer galaxy, where little or no HISA would otherwise be expected, because
the unperturbed line-of-sight velocity field is monotonic with distance.
With future SPH simulations that include
feedback mechanisms, we will discern whether stellar processing of interstellar gas leads to a closer match between the
model and actual galaxy.  We will also track the evolutionary trajectory of atomic clouds seen as HISA, to investigate
further the early stage of molecular cloud formation.


\section*{Acknowledgments}

We wish to thank the referee for making excellent suggestions which aided greatly in the presentation of this 
work.  The research leading to these results has received funding from the European Community's
Seventh Framework Programme under grant agreement n$^{\rm o}$ PIIF-GA-2008-221289.
C.L.D.'s research at Exeter was conducted as part of the award ``The formation of stars and planets:
Radiation hydrodynamical and magnetohydrodynamical simulations," made under the European Heads of Research 
Councils and European Science Foundation EURYI (European Young Investigator) Awards scheme, and supported 
by funds from the Participating Organisations of EURYI and the EC Sixth Framework Programme.
Calculations with {\Tor} were performed using the University of Exeter Supercomputer.
The research presented in this paper has used data from the Canadian Galactic Plane Survey, a Canadian 
project with international partners supported by the Natural Sciences and Engineering Research Council.


\bsp

\label{lastpage}

\end{document}